# Prediction of Ternary Fluorooxoborates with Coplanar Triangle Units $[BO_xF_{3-x}]^{x-}$ From First-Principles

Zhonglei Wei, [ab] Wenyao Zhang, [ab] Hao Zeng, [a] Hao Li, [ab] Zhihua Yang,[a*] Shilie Pan[a*]

*a.* CAS Key Laboratory of Functional Materials and Devices for Special Environments, Xinjiang Technical Institute of Physics & Chemistry, CAS;

*b.* Center of Materials Science and Optoelectronics Engineering, University of Chinese Academy of Sciences, Beijing 100049, China.

E-mail: zhyang@ms.xjb.ac.cn;
       slpan@ms.xjb.ac.cn.

Ten new ternary fluorooxoborate structures ($B_2O_2F_2$, I-X) were obtained from first-principles prediction. Coplanar aligned triangle structure units $[BO_2F]^{2-}$ and $[BOF_2]^-$ like $[BO_3]^{3-}$ in borates were found from the computational simulation. We identified new covalent coordination patterns of the F atom connected with the B atoms which are located in the bridging site, -B--F--B-. Besides, one molecular crystal with $[B_4O_4F_4]$ molecular unit was attached.

As a branch of inorganic crystal materials,[1-2] fluorooxoborates exhibit a diverse structural chemistry leading to fascinating properties and are used widely in many fields, such as ionic conductor for solid state batteries,[3] birefringent crystals for fiber optics isolators,[4-8] and nonlinear optical (NLO) frequency conversion material.[9-12] In particular, $NH_4B_4O_6F$,[10] $CsB_4O_6F$,[11] etc. are promising NLO materials for next generation deep-ultraviolet (DUV) all-solid-state laser devices.[12]

The general structure types of fluorooxoborates are as follow: A-site cations with oxyfluoride chromophores $[BO_xF_{4-x}]$ (x = 1, 2, 3), sometimes with $[BO_X]$ (x=3, 4) existed in the structure, A-site cations can be alkali metals, alkaline-earth metals, post-transition metals, such as $NaBOF_2$,[13] $SnB_2O_3F_2$,[9] $KCsB_8O_{12}F_2$,[14] $SrB_5O_7F_3$,[8] $BaBOF_3$,[15] etc. Borate systems also provide similar species of configuration.[16] However, among the borate and phosphate systems, binary ($B_2O_3$, $B_6O$, $P_2O_5$), ternary ($BPO_4$, $NPO_4$, $SiP_2O_7$), etc., there exist simple chemical components crystal structures without the A-site cations. We deployed a thorough investigation in the Inorganic Crystal Structure

Database (ICSD; FIZ Karlsruhe 2019, version 4.2.0), no trace of ternary fluorooxoborates ($B_xO_yF_{3x-2y}$, x, y=1, 2…) have been found. But, $C_2O_3F_2$,[17] $POF_3$ [18] crystallize in molecular crystal structure and $AlOF$[19] crystallizing in covalent crystal structure wer discovered, which gives us power and inspiration to study the ternary fluorooxoborates by computational analysis. Secondly, removing the A-site cations from crystal structures has become a viable proposition, $Be_2BO_3F$ is derived from the notable $KBe_2BO_3F_2$ obtained through experiment, and $Be_2BO_3F$ owns smaller interlayer space.[26] Besides, there are typical models which can be used for theoretical design, $BaBPO_5 \rightarrow BPO_4$,[27-28] $SrB_4O_7 \rightarrow B_2O_3$,[29-30] $K_2Al_2O_3F_2 \rightarrow AlOF$,[19, 31] etc. In the fluorooxoborate system, many new quaternary materials only with [$BO_xF_{4-x}$] anionic units were synthesized, $NaBOF_2$,[13] $KB_2O_3F$,[32] $RbBOF_2$,[33] $BaBOF_3$,[15] $SnB_2O_3F_2$[9], $PbB_2O_3F_2$,[9] with simple component ratio. Based on the above-mentioned factors, removing the A-site cations is a practicable strategy in the fluorooxoborate system.

With Materials Genome Initiative[20] and fast development of first-principles methods in researching the materials properties, high-throughput computations have emerged recently as a powerful way to search for materials rapidly with adequate properties among the large number of possible chemical compounds. In addition, molecular design strategies will be assisted to the development of structure prediction procedure.[21-26] Fluorooxoborates constitute an important class of materials with wide industrial applications, but ternary fluorooxoborates have been rarely investigated. It is of great interest to study these simpler composition systems, which may provide molecular-level understanding of the emergence of new properties and perception into designing new fluorooxoborates materials.

In this work, the prediction of ternary fluorooxoborates was put into effec with a simple molar ratio, B-O-F, 1:1:1 (the valence of B, O, F is +3, -2, -1, respectively), nine covalent crystal structures and one molecular crystal structure with the component of $B_2O_2F_2$ (Z=1,2,4,6) were predicted by using random structure search. Besides, the stability of dynamic, and thermodynamic analysis was performed. Furthermore, crystal structures, chemical bonding, and optical properties were carried out by the first-principles calculations.

In total, ten energetically favored structures in $B_2O_2F_2$ type (cell size, z=1, 2, 4, 6) were obtained. The DFT-relaxed lattice parameters, cell angles, atomic coordinates, are given in the ESI (Table S1). The basic requirement of the thermodynamic stability criteria is that the formation enthalpy $H_f < 0$, which ensures that the compound is stable against decomposition into the elements. We calculated the enthalpy of formation per atom using the following formula $H_f(B_2O_2F_2) = [E(B_2O_2F_2)-2E(B)-E(O_2)-E(F_2)] / (6 \times$ cell size), $E$ is the calculated enthalpy for each compound at ambient pressure. Our calculation indicates that the *Cc* phase with flat two-dimensional structure is energetically more stable than other phases, we take its enthalpy of formation as substrate. And the relative enthalpies of energy are shown in Figure 1a, the result meets stability criteria. The phase diagram of B-O-F ternary system is shown in Figure 1b which is generated with the help of Materials Project,[20] thermodynamically stable and unstable phase are represented by different colors, from Figure 1b, we can see that the title $B_2O_2F_2$ system occupies a central position. Furthermore, we checked the dynamical stability of the crystal structures by exploring phonon dispersions. As can be seen in Figure S2, no imaginary mode is observed for the ten predicted structures, confirming their dynamical stability.

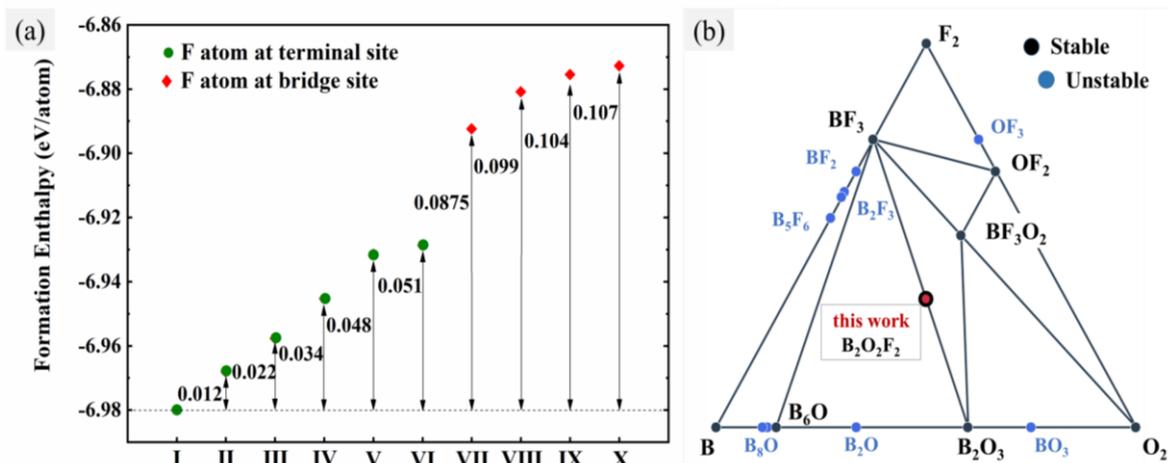

**Figure 1.** (a) Relative enthalpies of formation for different $B_2O_2F_2$ phases at ambient pressure. (b) Phase diagram of B-O-F system, phase diagram is generated with the help of Materials Project, https://materialsproject.org.

All the structures are listed in Figure 2, with their symmetries and structure units. $B_2O_2F_2$ I-X in this paper own various space groups: triclinic *P*-1, monoclinic *P*2, *P*2$_1$, *Cc* and *C*2/*c*, orthorhombic *P*222$_1$, *I*2$_1$2$_1$2$_1$, *Pmn*2$_1$ and *Pnma*. Until now, in the known architectures of fluorooxoborates, a fluorine atom usually connects directly with one B atom, and then connects with metal cation due to the highest electronegativity (Figure S3). However, as shown in Figure 2, we obtained two different covalent connected manners for fluorine atoms, i.e., terminal site F atoms connect with one B atom(-B--F) in structures I-VI, which is similar with the F atoms in the known fluorooxoborate structures, and bridging site F atoms (-B--F--B-) in structures VII-X, which is a new type. Meanwhile, the relative enthalpies of formation indicate that the crystal structures with terminal site F atoms have lower energy. Therefore, we need to conduct a comprehensive analysis of the structure and bonding mechanism. Here, we divided the structures into two groups in terms of the different coordination environment of F atoms, structures I-VI and VII-X, respectively. And these structures of the two types show different dimensional anionic architectures. Besides, we also obtained one molecular crystal structure (structure - IV) with 8-membered ring $<B_4$-$O_4>$, and it crystallizes in triclinic space group *P*-1, exhibiting a ring-layer structure, $<B_4O_4F_4>$ molecular unit puts four [$BO_2F$] together endways (Figure 2a).

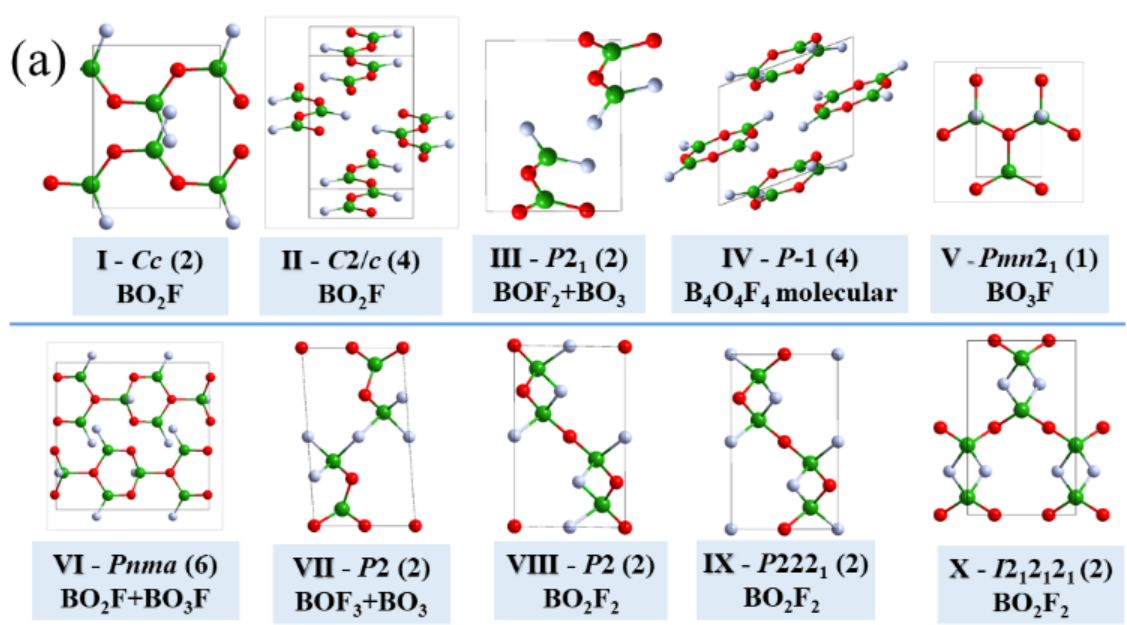

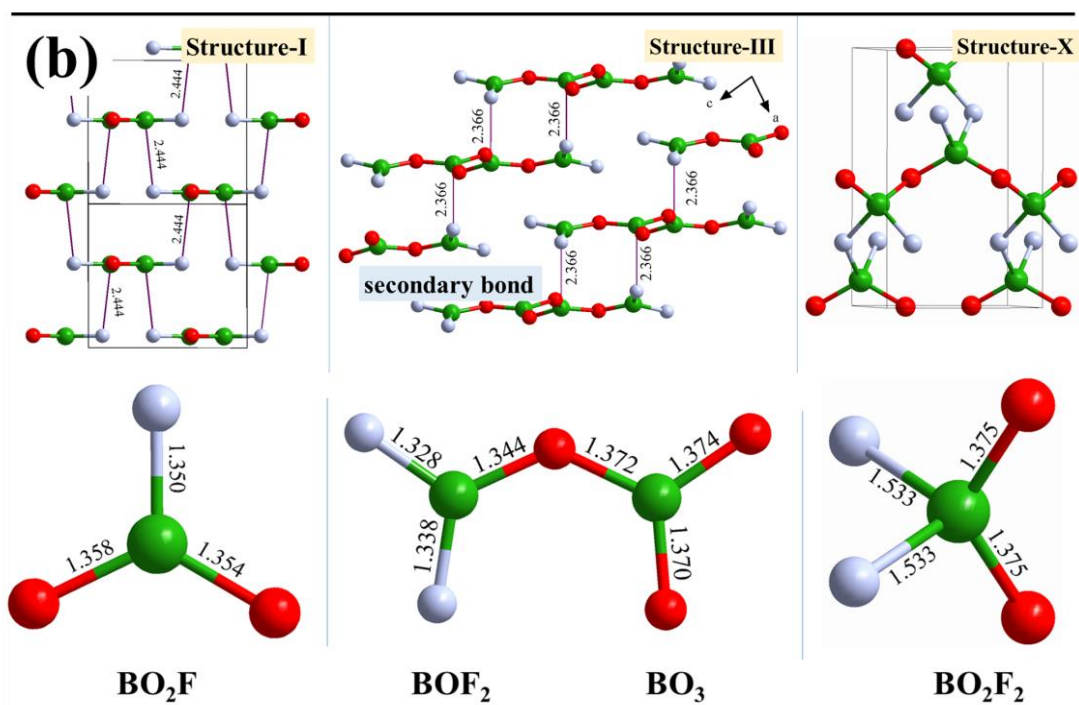

**Figure 2.** (a) Crystal structure of $B_2O_2F_2$ (I-X) with corresponding space group, structure units, and cell size. (b) Sketch map of I, III, X and their corresponding structure units; oxygen red, fluorine silvery, boron green.

Electronegativity[34-35] is one of the most important chemical descriptors for atoms. One important use of electronegativity is a predictive indicator for the formation of chemical bond. In 1932, Pauling et.al[36] discussed that if an atom has an adequately higher electronegativity than adjacent position, it could attract electron from its neighbor and oxidize the other atom, chemical bond thus comes into being (Figure S1). Moreover, in 1972, secondary bonding was first stated by Alcock that used to describe unconventional bonds among nonmetallic elements.[37] For the two types, concerning terminal site F and bridging site F of the 10 new structures, we would take a detailed crystal structure and chemical bonding analysis for exploring structure-property relationship. For brevity, we choose structures I, III, X in Figure 2b as main representative examples in the present discussion. In order to understand better chemical bond, we performed crystal orbital Hamilton population (COHP) analysis by using the LOBSTER code to study the bonding interaction.[38]

Figure S3 contains the important interatomic distances and Table S3 shows their integrated COHP data. Structure I crystallizes in the monoclinic system with the space group *Cc*, and it consists of the coplanar [BO$_2$F] unit, the [BO$_2$F] connects with each other by O atoms to form a [BO$_2$F]$_\infty$ chain, and further connect together via B-F secondary bonds between layers to form three-dimensional (3D) network. In this structure, the length of the B-F bond is 1.352 Å, and the B-O bond is 1.354-1.358 Å, which are similar with the bond length in conventional fluorooxoborates. Besides, the B-F secondary bond (2.444 Å) is shorter than van der Waals distances found (Figure 2b). III crystallizes in monoclinic crystal system with an acentric space group *P*2$_1$, its structure is composed of two-dimensional [BO$_3$ + BOF$_2$]$_\infty$ wavy layer, further bridged by the B-F secondary bonds (2.366 Å) to construct a 3D framework. The space group of structure X is orthorhombic system, *I*2$_1$2$_1$2$_1$, [BO$_2$F$_2$] is the exclusive building unit, and in this structure, all F atoms locate at a bridge site. The [BO$_2$F$_2$] unit is distorted with the lengthened B-F bond ~1.5 Å, much longer than ~1.3 Å in the terminal site F structures. It is derived from electrostatic interaction and the highest electronegativity of fluorine atom, and we will not take it as a secondary bond, because it is still a covalent interaction.

As can be seen in Figure S3 and Table S3, the energy overlap range and integrated COHP (ICOHP) value of conventional B-O and B-F bond are similar with each other among these three structures, but the ICOHP value of B-F bond 1.53 Å in structure X is much different, despite the fact that the bond length difference is only 0.2 Å. In addition, unlike van der Waals force, B-F secondary bonds are not purely an electrostatic attraction force, we can find obvious bonding and antibonding interaction owing to the weaker charge-transfer.

To the best of our knowledge, among the known fluorooxoborates, only tetrahedron elements [BO$_x$F$_{4-x}$] (x= 1, 2, 3) are found, but in structures I and III, the [BO$_2$F] and [BOF$_2$] units are coplanar aligned triangle as [BO$_3$] in borates. And in structures VIII-X, the [BO$_2$F$_2$] unit is an exclusive building unit, all F atoms are located in bridging site. Although they are obtained from computational simulation, it indicates that we should require much more effort to explore new experiment synthetic methods.

Fluorooxoborates attract increasing attention due to their excellent performance as

DUV NLO materials for the development of all solid-state laser. In principle, the prerequisites for a DUV NLO crystal should meet the following criteria: acentric crystal structure, wide bandgap ($E_g$>6.2 eV), large second harmonic generation (SHG) coefficient ($d_{ij}$ >0.39 pm/V) and moderate birefringence (>0.07@1064 nm), besides, easy growth of large-sized single crystals is also important for practical application.[39-40] Rremoval of A-site cations can be a more effective way to to enhance the DUV NLO performance from theoretical design.[41] There are many ternary materials treated as promising NLO materials, i.e., phosphorus nitrides $NaPN_2$ and polymer systems $PNF_2$, are predicted to be used as novel NLO frequency conversion materials in inorganic nitride systems.[42]

We took the predicted ternary fluorooxoborates structures as potential NLO materials and investigated their optical properties. The band gap, SHG coefficients and birefringence of all structures were calculated by using DFT method for simulation. With the help of PWmat,[42] the hybrid exchange correlation functional HSE06 was used to calculate band gap due to its good accuracy. As shown in Table 1, all the predicted band gaps of listed $B_2O_2F_2$ using HSE06 functional are larger than 8 eV, corresponding to a UV cutoff edge lower than 155 nm, which indicates the applicability of ternary fluorooxoborates in DUV region.

As can be seen in Table 1, structure III owns excellent DUV NLO properties, including a wide bandgap ($E_g \approx 9.7$ eV), colossal SHG coefficient ($d_{22} \approx 1.41$ pm/V, 3.6 x KDP, $d_{36}^{KDP}$=0.39 pm/V) and large birefringence ($\Delta n \approx 0.122$), it can realize much shorter DUV phase matching SHG output $\lambda_{PM} \approx 140$ nm. Structure I crystallizing in space group *Cc* with excellent UV NLO properties. As mentioned above, structures I and III own weak layering tendency due to small interlayer space (< 2.5 Å) and the existence of the B-F secondary bonds, which is shown in Figure S4.

**Table 1.** Crystallographic space Group, anionic motif, HSE06 fundamental gap (Eg), and linear and NLO properties of all the predicted structure $B_2O_2F_2$ (I-X).

|      | Space group | Anionic Motif | $E_g$ (eV) | $d_{ij}$ (pm/V) | $\Delta n$ (1064 nm) | $\lambda_{PM}$ (nm) |
|------|-------------|---------------|------------|-----------------|----------------------|---------------------|
| I    | $Cc$        | $BO_2F$       | 8.88       | $-1.78(d_{23})$ | 0.09                 | 212                 |
| II   | $C2/c$      | $BO_2F$       | 9.01       | /               | 0.125                |                     |
| III  | $P2_1$      | $BOF_2 + BO_3$ | 9.7       | $1.41(d_{22})$ $1.35(d_{23})$ | 0.122 | 140           |
| IV   | $P\text{-}1$ | $B_4O_4F_4$ molecular | 8.26 | /          |                      |                     |
| V    | $Pmn2_1$    | $BO_3F$       | 10.9       | $0.37(d_{33})$  | 0.017                |                     |
| VI   | $Pnma$      | $BO_2F+BO_3F$ | 8.95       | /               | 0.11                 |                     |
| VII  | $P2$        | $BOF_3 + BO_3$ | 8.5       | $0.85(d_{14})$  | 0.035                |                     |
| VIII | $P2$        | $BO_2F_2$     | 9.68       | $0.43(d_{14})$  | 0.003                |                     |
| IX   | $P222_1$    | $BO_2F_2$     | 9.54       | $0.63(d_{14})$  | 0.006                |                     |
| X    | $I2_12_12_1$ | $BO_2F_2$    | 9.44       | $0.8(d_{14})$   | 0.016                |                     |

Structures VIII, IX, X are unique bridging site F atom structures, $[BO_2F_2]$ is the only structure unit. With similar bandgaps, but optical properties are different. Among different structural symmetries $P2$, $P222_1$, $I2_12_12_1$, the largest second-order NLO coefficient $d_{14}$ is 0.43, 0.63, 0.8 pm/V and birefringence is 0.003, 0.006, 0.016, respectively. So the intermolecular arrangements and their microscale symmetries would influence the optical properties.[43]

In summary, ten ternary simple structures were obtained by first-principles prediction. Taking the research of feasibility at first, we computed structural, thermodynamic, and electronic properties of these structures. The basic linear and nonlinear optical properties of these crystals are reviewed which indicate that they satisfy performance requirement. Once confirmed by the following experiments, ternary fluorooxoborates would be a potential NLO material. Finally, it requires us to investigate new experiment synthetic methods to explore ternary structures and $[BO_2F]$ / $[BOF_2]$ coplanar structure units should be a top priority.

This work is supported by the National Natural Science Foundation of China (Grant

Nos. 51922014, 11774414, 51972336, 61835014), Basic Frontier Science Research Program of CAS (Grant No. ZDBS-LY-SLH035), Shanghai Cooperation Organization Science and Technology Partnership Program (Grant No. 2017E01013), Fujian Institute of Innovation, CAS (Grant No. FJCXY18010202).

**Conflicts of interest**

There are no conflicts to declare

# Electronic Supplementary Information

# Prediction of Ternary Fluorooxoborates with Coplanar Triangle Units [BO$_x$F$_{3-x}$]$^{x-}$ From First-Principles

Zhonglei Wei, [ab] Wenyao Zhang, [ab] Hao Zeng, [a] Hao Li, [ab] Zhihua Yang, [a*] and Shilie Pan [a*]

[a]CAS Key Laboratory of Functional Materials and Devices for Special Environments, Xinjiang Technical Institute of Physics and Chemistry, CAS; Xinjiang Key Laboratory of Electronic Information Materials and Devices, 40-1 South Beijing Road, Urumqi 830011, China.
[b]Center of Materials Science and Optoelectronics Engineering, University of Chinese Academy of Sciences, Beijing 100049, China.
E-mail: zhyang@ms.xjb.ac.cn
slpan@ms.xjb.ac.cn

# Contents

## Computational methods

The first-principles calculations are performed by the psuedopotential methods implemented in the CASTEP package [1--2] based on the density functional theory (DFT). The optimized norm-conserving pseudopotentials [3-4] are used to simulate ion-electron interactions for all constituent elements. The exchange-correlation functional is approximated using the generalized gradient approximation (GGA) as parametrized by Perdew–Burke–Ernzerhof (PBE).[5] Kinetic energy cutoff of 850 eV is chosen with Monkhorst-Pack k-point meshes spanning 0.04/Å$^3$ in the Brillouin zone. The cell parameters and the atomic positions in the unit cells of all crystals are fully optimized using BFGS method. The phonon spectrum of the stable compound was calculated using the linear response method. The convergence thresholds between optimization cycles for energy change, maximum force, maximum stress, and maximum displacement are set as $5.0\times10^{-6}$ eV/atom, 0.01 eV/Å, 0.02 GPa, and $5.0\times10^{-4}$ Å, respectively. And the hybrid HSE06 functional implemented in Pwmat code [6a] was adopted for more accurate bandgap value, HSE_alpha default value 0.25 and NCPP-SG15-PBE pseudopotential was used in all calculations.[6b] The empty bands were set as 3 times that of valence bands in the calculation to ensure the convergence of optical properties.

Structure search method in this paper is implemented within Random Searching Algorithm which was presented in our group, certain crystal cell parameter and elements are given for structures. The final structure is derived from natural transition of energy evolution, and we will give a summary for this method in the future. Just like the known crystal structure prediction software USPEX and CALYPSO, the period of structure prediction is long and needs a large quantity of computing resources. In this paper, our goal is not for structure with minimum potential energies, but to explore ternary fluorooxoborates system (B-O-F). In the end, we make the effort to explore minimum potential energies structure within $B_2O_2F_2$ (Z=2), and to our best knowledge, the result indicates that the structure-I *Cc* phase in this paper is the lowest energy structure within using genetic algorithm (GA), which is an excellent way in the crystal structure prediction software USPEX, [7-8] and have done our best, there are no more innovative stable structures.

**Table S1.** The crystallographic data of $B_2O_2F_2$ (I~X) in prediction.

| $B_2O_2F_2$ - I ( Z=2 ) | | | | | | |
|---|---|---|---|---|---|---|
| *Cc* (9) | | | | | | |
| Unit cell parameters | a (Å) 5.6081 | b (Å) 5.3215 | c (Å) c=8.5860 | α (°) 90 | β (°) 154.1932 | γ(°) 90 |
| | | | | | | |
| Fractional coordinates | x/a | | y/b | | z/c | |
| B | 0.64296 | | -0.35830 | | 0.56202 | |
| O | 0.83325 | | -0.15181 | | 0.74865 | |
| F | 1.27379 | | -0.09081 | | 0.68934 | |

| $B_2O_2F_2$ - II ( Z=4 ) | | | | | | |
|---|---|---|---|---|---|---|
| *C2/c* (15) | | | | | | |
| Unit cell parameters | a (Å) 10.1411 | b (Å) 5.2225 | c (Å) 4.3038 | α (°) 90 | β (°) 74.3269 | γ(°) 90 |
| | | | | | | |
| Fractional coordinates | x/a | | y/b | | z/c | |
| B | 0.36967 | | 0.11689 | | 1.15859 | |
| O | 0.37022 | | 0.09699 | | 1.47509 | |
| F | 0.37409 | | 0.35067 | | 1.03428 | |

| $B_2O_2F_2$ - III ( Z=2 ) | | | | | | |
|---|---|---|---|---|---|---|
| $P2_1$ (4) | | | | | | |
| Unit cell parameters | a (Å) 5.4880 | b (Å) 4.2425 | c (Å) 4.7742 | α (°) 90 | β (°) 79.0899° | γ(°) 90 |
| | | | | | | |
| Fractional coordinates | x/a | | y/b | | z/c | |
| B1 | 0.31806 | | 0.95486 | | 0.39300 | |
| B2 | 0.93242 | | 0.43226 | | 0.12370 | |
| O1 | 0.77741 | | 0.31242 | | 0.35891 | |
| O2 | 0.00381 | | 0.24481 | | 0.88773 | |
| F1 | 0.73459 | | 0.75188 | | 0.66840 | |
| F2 | 0.48203 | | 0.80377 | | 0.19907 | |

| B₂O₂F₂ - IV    molecular crystal (Z=4) | | | | | |
|---|---|---|---|---|---|
| *P*-1    (2) | | | | | |
| Unit cell parameters | a (Å) | b (Å) | c (Å) | α (°) | β (°) | γ(°) |
| | 5.5026 | 6.5293 | 6.7834 | 87.1609 | 70.6584 | 97.3439 |

| Fractional coordinates | x/a | y/b | z/c |
|---|---|---|---|
| B | 0.00452 | 0.13928 | 0.25984 |
| B | 0.55266 | 0.36451 | 0.21553 |
| B | 0.55859 | 0.28365 | 0.84967 |
| B | 0.07691 | 0.22021 | 0.60654 |
| O | 0.89561 | 0.93961 | 0.27219 |
| O | 0.50194 | 0.55844 | 0.25037 |
| O | 0.57339 | 0.26449 | 0.04336 |
| O | 0.10080 | 0.22380 | 0.40155 |
| F | 0.05466 | 0.25836 | 0.08153 |
| F | 0.38707 | 0.88008 | 0.26014 |
| F | 0.96480 | 0.60304 | 0.30671 |
| F | 0.60117 | 0.26017 | 0.37053 |

| B₂O₂F₂ - V    (Z=1) | | | | | |
|---|---|---|---|---|---|
| *Pmn*2₁    (31) | | | | | |
| Unit cell parameters | a (Å) | b (Å) | c (Å) | α (°) | β (°) | γ(°) |
| | 2.4768 | 4.4132 | 4.1500 | 90 | 90 | 90 |

| Fractional coordinates | x/a | y/b | z/c |
|---|---|---|---|
| B | 0.50000 | 0.62294 | 0.04867 |
| O | 0.50000 | 0.49320 | 0.38801 |
| F | 1.00000 | 0.07376 | 0.56332 |

| B₂O₂F₂ - VI    (Z=6) ||||||
|---|---|---|---|---|---|
| *Pnma*   (62) ||||||
| Unit cell parameters | a (Å) | b (Å) | c (Å) | α (°) | β (°) | γ(°) |
| | 8.1549 | 7.8786 | 4.8907 | 90 | 90 | 90 |
| Fractional coordinates | x/a || y/b || z/c ||
| B | 0.04820 || 0.25000 || 0.39321 ||
| B | 0.33817 || 0.40466 || 0.33102 ||
| O | 0.24516 || 0.25000 || 0.36255 ||
| O | 0.49089 || 0.39973 || 0.24264 ||
| F | 0.51712 || 0.25000 || 0.83066 ||
| F | 0.26416 || 0.54798 || 0.40681 ||

| B₂O₂F₂ - VII    (Z=2) ||||||
|---|---|---|---|---|---|
| *P*2   (3) ||||||
| Unit cell parameters | a (Å) | b (Å) | c (Å) | α (°) | β (°) | γ(°) |
| | 6.9489 | 3.9955 | 3.9225 | 90 | 94.0939 | 90 |
| Fractional coordinates | x/a || y/b || z/c ||
| B | 0.09372 || 0.09145 || 0.28673 ||
| B | 0.63730 || 0.46215 || 0.76543 ||
| O | 0.00000 || 0.21066 || 0.00000 ||
| O | 0.00000 || 0.88432 || 0.50000 ||
| O | 0.73530 || 0.23325 || 0.59353 ||
| F | 0.50000 || 0.28668 || 0.00000 ||
| F | 0.50000 || 0.63889 || 0.50000 ||
| F | 0.27806 || 0.70287 || 0.03819 ||

| B₂O₂F₂ - VIII (Z=2) |||||||
|---|---|---|---|---|---|---|
| *P*2   (3) |||||||
| Unit cell parameters | a (Å) | b (Å) | c (Å) | α (°) | β (°) | γ(°) |
| | 6.5281 | 3.9617 | 3.9745 | 90 | 90.8235 | 90 |
| | | | | | | |
| Fractional coordinates | x/a || y/b || z/c ||
| B | 0.62405 || 0.46319 || 0.26030 ||
| B | 0.12665 || 0.01006 || 0.79246 ||
| O | 1.00000 || 0.83310 || 1.00000 ||
| O | 0.50000 || 0.33480 || 0.50000 ||
| O | 0.74921 || 0.25841 || 0.07914 ||
| F | 0.50000 || 0.67604 || 1.00000 ||
| F | 0.00000 || 0.17315 || 0.50000 ||
| F | 0.24979 || 0.75978 || 0.58154 ||

| B₂O₂F₂ - IX (Z=2) |||||||
|---|---|---|---|---|---|---|
| *P*222₁   (17) |||||||
| Unit cell parameters | a (Å) | b (Å) | c (Å) | α (°) | β (°) | γ(°) |
| | 2.4768 | 4.4132 | 4.1500 | 90 | 90 | 90 |
| | | | | | | |
| Fractional coordinates | x/a || y/b || z/c ||
| B | 0.28932 || 0.74269 || 1.12382 ||
| O | 0.50000 || 0.07794 || 0.75000 ||
| O | 0.42043 || 0.50000 || 1.00000 ||
| F | 1.00000 || 0.58191 || 1.25000 ||
| F | 0.92048 || 1.00000 || 1.50000 ||

| B₂O₂F₂ - X (Z=2) |||||||
|---|---|---|---|---|---|---|
| *I*2₁2₁2₁   (24) |||||||
| Unit cell parameters | a (Å) | b (Å) | c (Å) | α (°) | β (°) | γ(°) |
| | 4.0229 | 3.8929 | 6.5359 | 90 | 90 | 90 |
| | | | | | | |
| Fractional coordinates | x/a || y/b || z/c ||
| B | 0.50000 || 0.75000 || 0.39973 ||
| O | 0.25000 || 0.57791 || 0.50000 ||
| F | 0.16570 || 1.00000 || 0.75000 ||

**Table S2.** A summary of ICSD collection codes for the known ternary oxyfluoride materials $M_xO_yF_z$ among III A to VII A.

| Compounds/ (ICSD Collection Code) | Other type crystal | Molecular crystal |
|---|---|---|
| [BOF] | / | / |
| [AlOF] | 36556;   260261 | |
| [SiOF] | 159577;   171499 | |
| [GaOF] | / | / |
| [GeOF] | 250513 | |
| [AsOF] | / | / |
| [InOF] | 2521 | |
| [SnOF] | 948; 78356; 409393 | |
| [SbOF] | 21099;62476;67157; 78849;431207;431209; | |
| [TeOF] | 1275; 2173; 2174; 82162;88415 | |
| [COF] | | 163001; 249217 401780; 414408 |
| [NOF] | | 411510 |
| [POF] | | 63246; 248122;250498 |
| [SOF] | | 48148; 62968;66514 |
| [ClOF] | | 412426 |
| [SeOF] | | 12110 |
| [BrOF] | | 50199 |
| [IOF] | | 4076;201202;280804 |

**Table S3.** The ICOHP values (in eV/bond) are listed to show the corresponding interactions in Figure S3.

|  | I | | III | | X | |
|---|---|---|---|---|---|---|
|  | Length | ICOHP | Length | ICOHP | Length | ICOHP |
| B-O | 1.349 | -12.15 | 1.374 | -11.92 | 1.368 | -11.88 |
|  | 1.354 | -12.39 | 1.372 | -12.46 |  |  |
|  |  |  | 1.370 | -11.57 |  |  |
|  |  |  | 1.344 | -11.37 |  |  |
| B-F | 1.342 | -12.76 | 1.328 | -12.79 | 1.53 | -7.82 |
|  | 2.444 | -0.74 | 1.338 | -11.87 |  |  |
|  |  |  | 2.366 | -0.87 |  |  |
| O-F | 2.279 | -0.236 | 2.313 | -0.13 | 2.33 | -0.134 |
|  | 2.337 | -0.216 | 2.373 | -0.11 | 2.35 | -0.1 |

**Figure S1. Statistical and characteristic analysis of ternary oxyfluoride materials** (a) Part of the periodic table of elements; (b) Statistical analysis of the known ternary oxyfluoride materials among III A to VII A; (c) A representative covalent crystal AlOF; (d) A representative molecular crystal $P_2O_3F_4$; (e) A representative which indicates that electronegativity difference is an indicator for chemical bond, $ClOF_3$. (oxygen red, fluorine silvery)

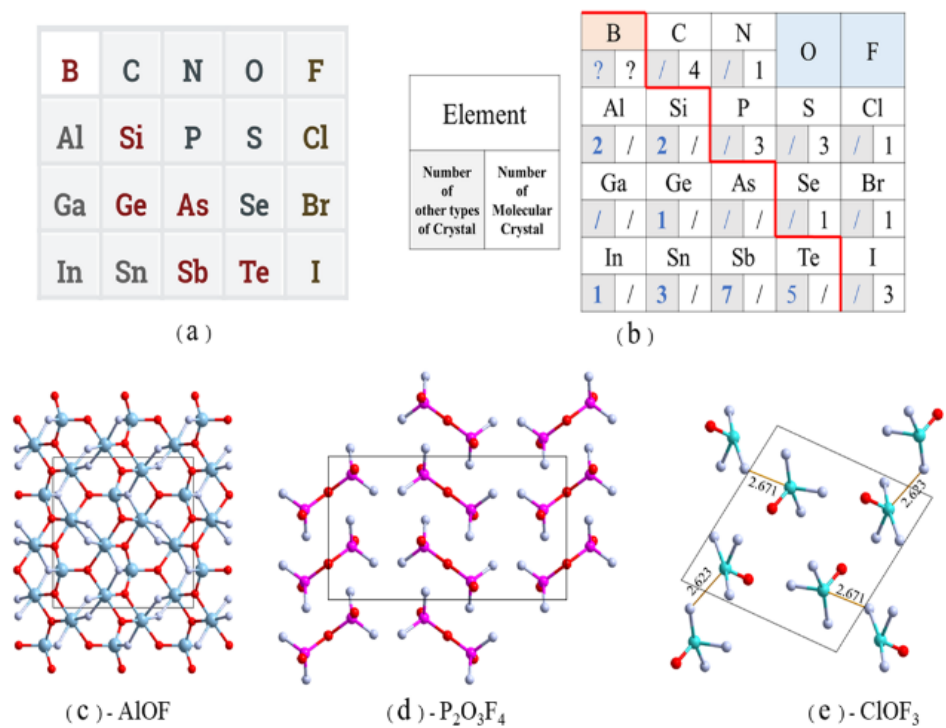

Periodic Table of the chemical elements is a powerful facility for exploring the property of elements, what it found especially interesting is its elegance and predictive power. In the Periodic Table, adjacent elements must be closely related to their physical and chemical functions. As can be seen in Figure S1(a), boron element (B) is the first metalloid element from top to bottom on the Periodic Table, transition metals and nonmetallic elements are screened off from metalloid element. Figure S1(b) shows further statistical analysis performed on the type of crystal among the known ternary oxyfluoride materials $M_xO_yF_z$ (M is element in main group III A - VII A) in the available materials database, ICSD 2019 - v4.2.0. It could be found that when we take metalloid elements as the dividing line, to date, molecular structure crystals only exist in the right side, and corresponding ICSD collection codes are listed in Table S2. Even so, a molecular crystal in the prediction of ternary system [B-O-F] was obtained here, it needs further experimental verification.

Oxyfluoride $ClOF_3$,[9] as shown in Figure S1(e), the Cl element, general negative valence (-1, CsCl), acts as a central cation (Cl, +5) due to the smallest electronegative value among Cl, O, F: 3.16, 3.44, 3.98 in Pauling scale, respectively.

**Figure S2.** The phonon spectra of $B_2O_2F_2$ (I-X).

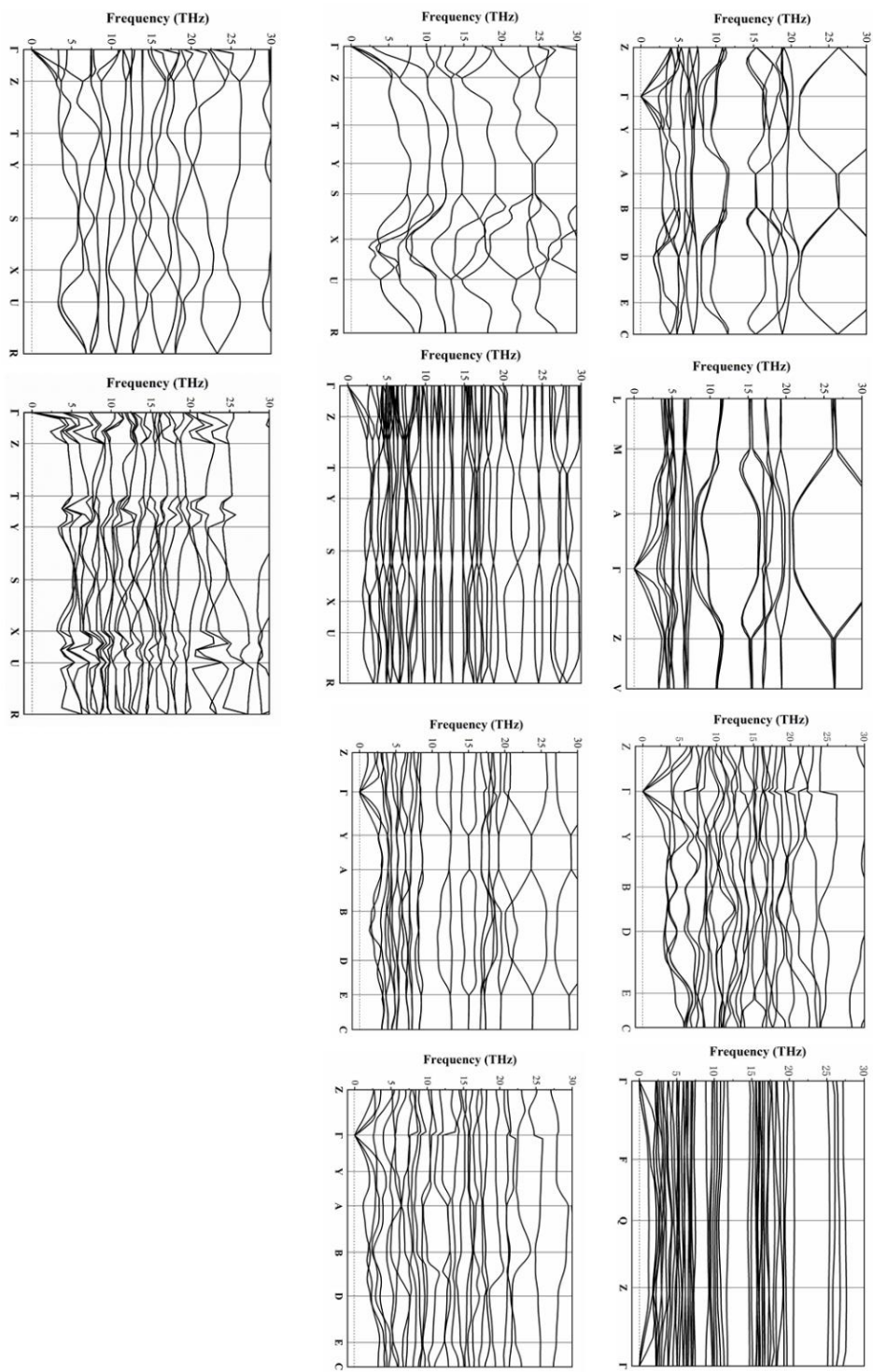

**Figure S3.** The crystal orbital Hamilton population (COHP) analysis of structure I, III, X bonding interactions based on DFT plane-wave calculations.

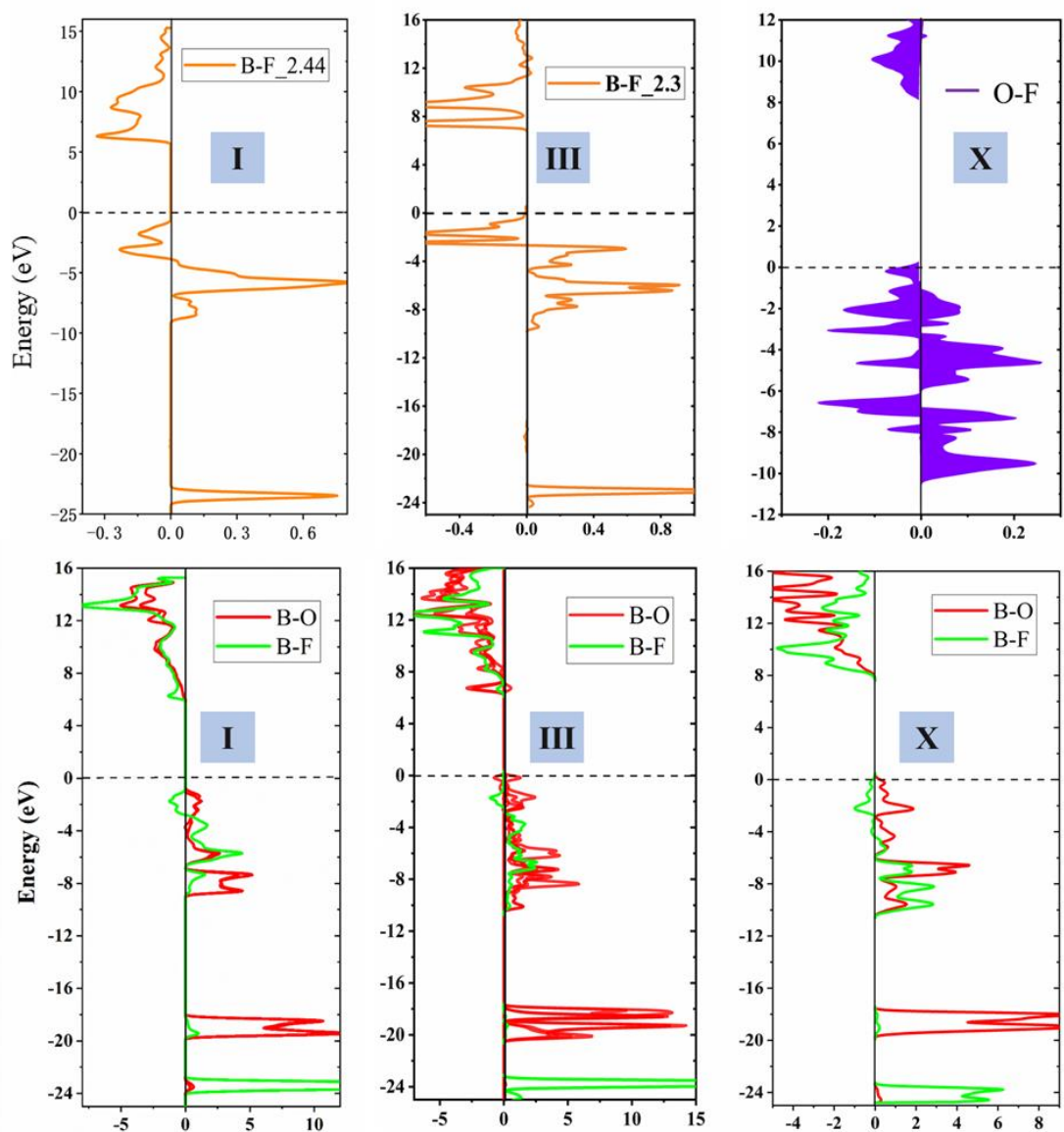

Negative value indicates antibonding interaction and positive value indicates bonding interaction.

**Figure S4.** The coordination architectures of fluorine atom (F) in the known structures and predicted structures.

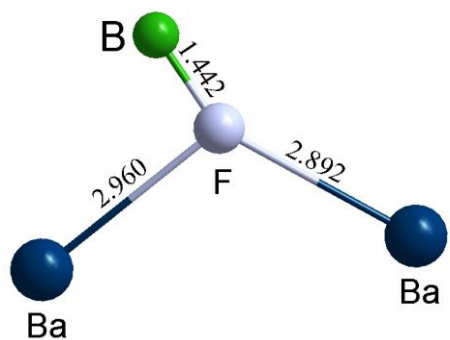

Bond length in conventional fluorooxoborates, B-F covalent interaction and Ba-F ionic interaction. Partial structure sketch map of alkaline-earth metal fluorooxoborate $BaB_4O_6F_2$.[10]

This structural characteristic is consistent with those of alkali-metal and transition metal fluorooxoborates.

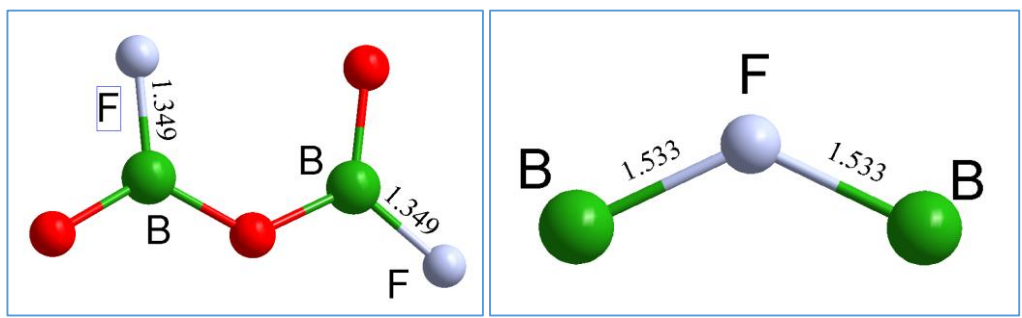

New covalent coordination pattern of F atom connected with the B atoms:
(1) terminal site F atoms (-B--F);
(2) bridging site F atoms (-B--F--B-).

**Figure S5.** The interlayer spacing of the structure I, III and KBe$_2$BO$_3$F$_2$ structure.

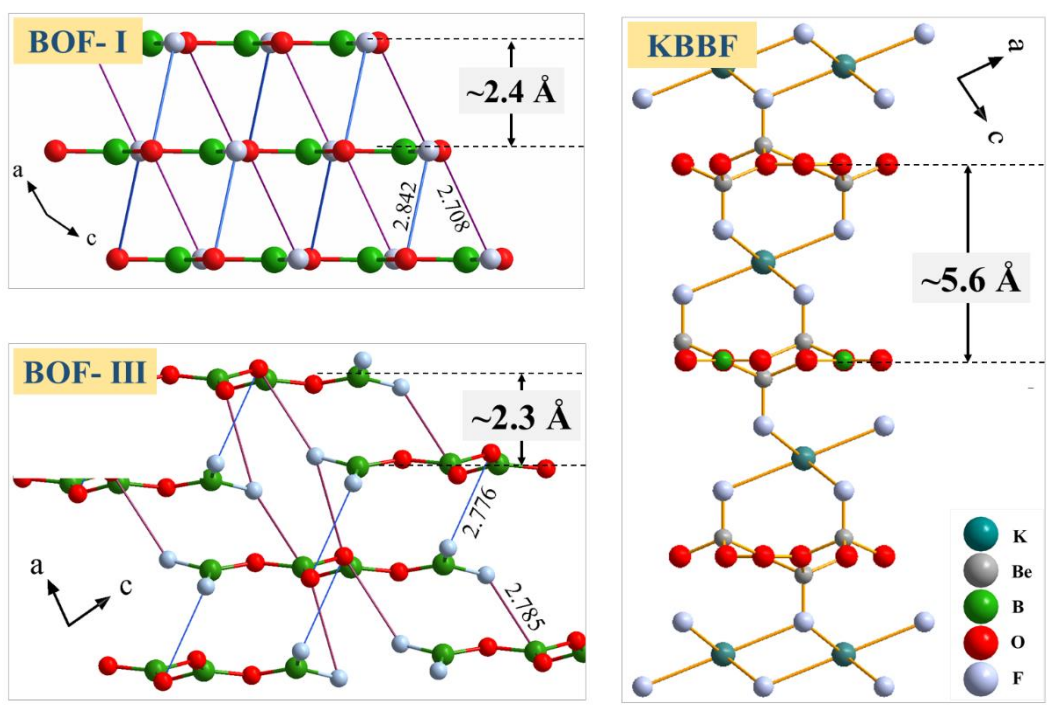

Predicted structure I, III with small interlayer space and B-F secondary bonds can effectively suppress the layering tendency.